\documentclass[aps,english,prb,twocolumn,showpacs,superscriptaddress]{revtex4}
\usepackage{graphicx}

\begin{document}

\title{Epitaxial growth mechanisms of graphene and effects of substrates}

\author{V. Ongun \" Oz\c celik}
\affiliation{UNAM-National Nanotechnology Research Center, Bilkent University, 06800 Ankara, Turkey}
\affiliation{Institute of Materials Science and Nanotechnology, Bilkent University, Ankara 06800, Turkey}

\author{S. Cahangirov}
\affiliation{UNAM-National Nanotechnology Research Center, Bilkent University, 06800 Ankara, Turkey}
\affiliation{Institute of Materials Science and Nanotechnology, Bilkent University, Ankara 06800, Turkey}

\author{S. Ciraci}\email{ciraci@fen.bilkent.edu.tr}
\affiliation{UNAM-National Nanotechnology Research Center, Bilkent University, 06800 Ankara, Turkey}
\affiliation{Institute of Materials Science and Nanotechnology, Bilkent University, Ankara 06800, Turkey}
\affiliation{Department of Physics, Bilkent University, Ankara 06800, Turkey}

\begin{abstract}

The growth process of single layer graphene with and without substrate is investigated using ab-initio, finite temperature molecular dynamic calculations within density functional theory. An understanding of the epitaxial graphene growth mechanisms in the atomic level is provided by exploring the transient stages which occur at the growing edges of graphene. These stages are formation and collapse of large carbon rings together with the formation and healing of Stone-Wales like pentagon-heptagon defects. The activation barriers for the healing of these growth induced defects on various substrates are calculated using the climbing image nudge elastic band method and compared with that of the Stone-Wales defect. It is found that the healing of pentagon-heptagon defects occurring near the edge in the course of growth is much easier than that of Stone-Wales defect. The role of the substrate in the epitaxial growth and in the healing of defects are also investigated in detail, along with the effects of using carbon dimers as the building blocks of graphene growth.

\end{abstract}

 \maketitle

\section{Introduction}

Unusual chemical and physical properties, such as high chemical stability, high carrier mobility and high mechanical strength have made graphene\cite{novoselov2004} an attractive material in fundamental and applied science. Pristine graphene, in particular, appears to be an important material to be used in future high-technology applications.\cite{lee2009,mehmet} In this respect, the production of epitaxial graphene has been the motivation of recent experimental and theoretical studies.\cite{berger2006}

The recent studies on graphene production processes can be grouped in two classes: one class includes the mechanically exfoliated graphene sheets where the graphene flakes are peeled from a bulk graphite substrate.\cite{novoselov2004, zhang20051, geim2007} However, that method has the disadvantage of not being able to easily control the size and quality of the fabricated layer. The other class includes the direct growth of graphene flakes on substrates.\cite{zhang20052, bunch2005} It was shown that epitaxial graphene is a suitable material for nanoscale electronic applications by growing ultrathin graphite layer on silicon carbide by thermal decomposition.\cite{berger2004}

There are two main events happening during epitaxial graphene formation: Nucleation and growth of graphene from the nucleated seed. Nucleation on a substrate is mostly favored by defects which are actually step edges in the atomistic scale. The nucleation process of carbon monomers and dimers on various transition metal substrates were investigated in a recent study to distinguish the substrates depending on monomer- or dimer-based carbon nucleation.\cite{chen2010} Mechanisms and factors influencing graphene growth were also investigated using Ru(0001), Pt(111), Ir(111) and Pd(111) substrates.\cite{sutter2008,sutter2009,loginova2009,murata2010} It was found that the orientation of grown graphene sheets and carbon concentration at their edge play crucial role in the growth process.\cite{loginova2009} Graphene was also grown on SiC substrates using high temperature sublimation.\cite{johansson2011} In a recent study, the stability of graphene on nickel surfaces and the healing of defects were investigated using Monte Carlo simulations with tight-binding potentials.\cite{karoui2010} Defected graphene flakes containing pentagons, heptagons, octagon and nonagons were treated at different temperatures and the healing effect of increasing temperature was observed. Also in this same study, nickel was proposed as a substrate for better defect healing at low temperature, which is consistent with the results of our calculations.

In this study, using ab-initio finite temperature molecular dynamics (MD) method, we investigate the atomistic mechanisms taking place during graphene growth by considering carbon atoms and carbon dimers as the building blocks. To clarify the effects of substrates, we consider the hypothetical growth of graphene without a substrate, as well as the growth on layered BN substrate.\cite{nickel}

Our calculations reveal and explain two major mechanisms observed during graphene growth. The first one is the formation of large carbon rings at the edges of the growing structure. With the inclusion of new carbon atoms, these large rings further expand and eventually collapse into smaller structures when some critical ring size is reached. The smaller structures formed after the collapse are composed of hexagonal and defected regions of graphene. The defected regions, which can form both before and after the collapse of rings, generally consist of pentagons and heptagons. Similar pentagon-heptagon structures were recently observed in grain boundaries of graphene grown by chemical vapor deposition.\cite{huang2011}

The second major mechanism of the growth process is the healing of these defected regions formed at the edges of the growing structure. In this respect, we investigate the healing process of defects composed of neighboring pentagons and heptagons which are named as pentagon-heptagon (PH) defects throughout the manuscript. PH defect is similar to the well-known Stone-Wales(SW) defect,\cite{stone86, wales98} but here some carbon atoms are two-fold coordinated with unsaturated $sp^{2}$-type bonds. The energy barrier for the healing of a PH defect is lower as compared to that of SW defect due to this deficiency. We also present the analysis for the energetics of healing of SW and PH defects in free standing graphene, as well as graphene grown on BN and Ni(111) substrates using the climbing image nudged elastic band (NEB) method.\cite{neb} We found that the healing of PH defects is further facilitated when the lattice of graphene grown on Ni(111) substrate is expanded.

\section{Method}

We have performed atomic structure optimizations and ab-initio finite temperature molecular dynamics (MD) calculations within density functional theory (DFT) using VASP software.\cite{kresse96}  The relaxed geometries of all structures were calculated by spin-polarized plane-wave calculations using projector augmented-wave (PAW) potentials\cite{blochl94} within generalized gradient approximation (GGA)\cite{perdew92} including van der Waals corrections.\cite{grimme2006} The Brilliouin zone of the primitive unitcell of graphene was sampled by (17$\times$17$\times$1) k-points in Monkhorst-Pack scheme which was scaled according to the size of other unitcells.\cite{monkhorst76} The energy convergence value between two consecutive steps was chosen to be $10^{-5}$~eV. In ab-initio MD calculations the time step was taken 2.5 fs and the atomic velocities were renormalized to the temperature set at T=1300 K at every 40 time steps. The temperature of MD calculations is in compliance with the temperature used in chemical vapor deposition (CVD).

\section{Molecular Dynamics Simulations of Growth}

\begin{figure}
\includegraphics[width=8cm]{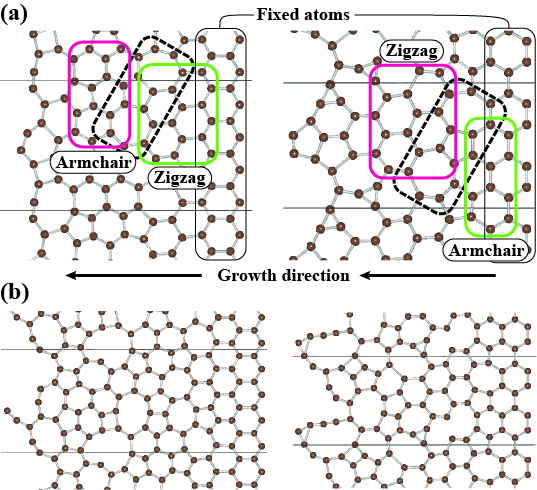}
\caption{Snapshots from ab-initio MD simulations of planar graphene growth at T=1300 K without a template substrate. An initial flake was placed and in each 1~ps MD calculation two carbon atoms were sent from the left hand side to monitor the growth in the indicated direction. Each snapshot includes two periodic supercells. (a) Change of the armchair edge to zigzag edge and vice versa is shown. (b) Structures obtained when simulation of growth presented in (a) is proceeded. Formation of big rings and chains were observed, and the resulting structures were far away from being a perfect graphene layer. Note that defects formed in part (a) are still present in part (b).}

\label{fig1}
\end{figure}

\begin{figure*}
\includegraphics [width=15cm] {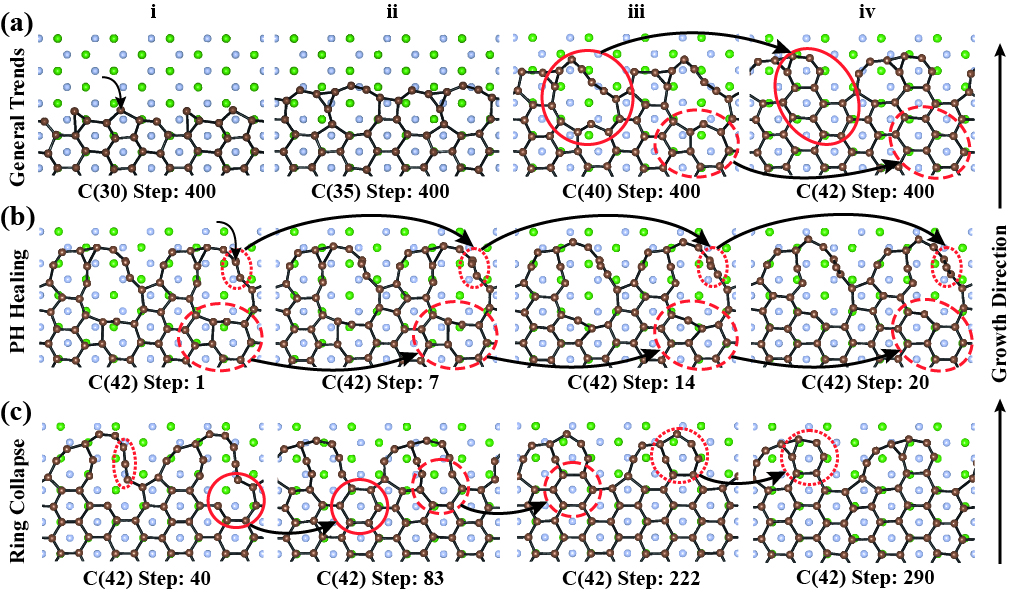}
\caption{Snapshots from ab-initio MD simulation of epitaxial growth of graphene on a BN substrate. In the ball and stick model B, N and C atoms are represented by green, blue and brown balls while only bonds between carbon atoms having distance less than 2~\AA~are shown. Each snapshot includes two periodic supercells in the horizontal direction. (a) General trends are presented by including final configurations of MD calculations involving 30, 35, 40 and 42 carbon atoms. Some of the critical configurations in the evolution of ring collapse and defect healing mechanisms are highlighted by solid and dashed lines respectively. (b) Snapshots from the MD simulation of the structure having 42 carbon atoms taken after 1, 7, 14 and 20 MD steps. Carbon atom migration causing the growth of rings and defect healing can be traced in dotted and dashed circles, respectively. (c) Snapshots from the same MD simulation taken after 40, 83, 222 and 290 MD steps. Three subsequent hexagon formations are indicated by solid, dashed and dotted circles.}

\label{fig2}
\end{figure*}

\subsection{Growth without Template}

In order to understand the effects of a template surface on graphene growth, we first consider a hypothetical situation and investigated the growth mechanism of graphene without any template (substrate). To bypass the initial nucleation process, a graphene flake was fixed in space and additional carbon atoms were sent to it in different scenarios, which mimic the growth. Simulations were done by letting the carbon atoms to move freely in a certain plane, while not allowing the out of plane motions. As shown in Fig.~\ref{fig1}(a), we start with nanoribbons having armchair and zigzag edges. Initially, zigzag nanoribbon is composed of 24 atoms forming three zigzag chains in the periodic direction, two of which are kept fixed. Armchair nanoribbon starts with 20 atoms, 12 of which are kept fixed. Fixed atoms are delineated in Fig.~\ref{fig1}(a). After running 1~ps of MD simulation two more carbon atoms are introduced in both systems. These atoms are first positioned in the same plane to the left of the nanoribbons, where the edges are free to move in 2D. Then they are moved towards these edges until the distance between the newcoming atom and one of the edge atoms decreases to 1.3~\AA. Then the new MD simulation is started and the process is repeated consecutively.

As seen in Fig.~\ref{fig1}(a), during the growth simulation the orientation of the honeycomb parts are changed from zigzag to armchair (left panel) and vice versa (right panel). Interestingly, in both cases the transition is mediated by similar structures composed of two heptagons with one pentagon in the middle. One can attribute the defected growth to the absence of a substrate which would act as a stencil if the template had a structure similar to graphene. Carrying on the growth simulation of structures presented in Fig.~\ref{fig1}(a) results in the massively defected network of carbon atoms as shown in Fig.~\ref{fig1}(b). One can identify the big holes surrounded by carbon chains and patterns composed of pentagons and heptagons. It was observed that the defects that emerged at the beginning of the simulation are still present after about 40~ps of simulation. This implies that the process of growth induces defects mainly composed of pentagons and heptagons, which persist due to the absence of a healing mechanism. Several scenarios of growth simulation without a substrate were tested but none of them resulted in a reasonably ordered honeycomb structures. Especially, those simulations in which carbon atoms were allowed to move in all directions resulted in bulk-like structures and atomic chains. Although the observed chain structures are interesting, growth of regular honeycomb structure was not observed and we deduced the necessity of a template during the growth process to define a plane where the graphene like structure sits and where the newcoming carbon atoms are landed.

\subsection{Dynamics of Graphene Growth on BN Substrate}

\subsubsection{Monomers}
Hexagonal boron-nitride consists of single layers of BN in honeycomb structure, which is almost commensurate to graphene. It has been argued that BN layers of any thickness can be grown on graphene layers and vise versa.\cite{liu11} Because of this reason we have chosen BN as a template on which we investigate the growth of graphene. Again, to skip the initial nucleation process, a graphene flake was placed on BN substrate. Single carbon atoms were released from random positions on top of the graphene flake edges and molecular dynamics simulation were performed for 400 time steps before the next atom was sent. By sending the atoms one by one, events happening during the growth process were monitored at atomistic scale. The snapshots taken from this calculation are shown in Fig.~\ref{fig2}. The bottom part of these structures normally comprises fixed graphene and BN substrate, which are not shown while growth proceeds upwards.

In Fig.~\ref{fig2}(a) we present the general trends observed during growth. Each of the four snapshots in this row corresponds to the final structures obtained after the MD simulations of 30, 35, 40 and 42 carbon atoms on BN respectively. As seen in the first column of Fig.~\ref{fig2}(a) first a single carbon atom indicated by a small arrow makes bonds with armchair edges and a pentagon structure is formed. This stretches the edges and prepares a medium for the formation of a neighboring heptagon. Then ring-like structures start to grow at the edges as seen in column two of Fig.~\ref{fig2}(a). When the ring structure reaches a certain size, it collapses forming hexagonal structures at the graphene edges as shown in columns three and four.

The formation and healing of pentagon-heptagon defects, which is the second major mechanism affecting the growth process is presented in Fig.~\ref{fig2}(b). The snapshots correspond to the 1$^{st}$, 7$^{th}$, 14$^{th}$ and 20$^{th}$ MD steps of the simulation with 42 carbon atoms. The healing of the PH defect is highlighted by dashed circles. As simulation proceeds to the 20$^{th}$ step, the PH defect is totally relaxed into two hexagons. Note that, the healing of a PH defect at the edge is similar to the healing of a SW defect, which involves rotation of the middle bond (which is common to two adjacent heptagons) by 90$^{\circ}$. Here there is, however, a crucial difference in the path of healing as compared to that of SW healing, because one of the carbon atoms in the pentagon of the PH defect is bonded with two adjacent carbon atoms, whereas in the SW defect all carbons are bonded to three others. The absence of one of these bonds (or the presence of $sp^{2}$-type dangling bond) decreases the barrier of PH healing as compared to the SW case. This issue is revisited in forthcoming detailed discussions. We note the growing edge of grains is reminiscent of the grain boundary.\cite{huang2011} The contact of an adjacent grain to the growing edge of graphene is expected to delay the healing of defects.

In Fig.~\ref{fig2}(b), the dotted circle in the first column marks the inclusion of the newly added carbon atom which is added from a random position on top of the graphene layer. This newly added carbon atom is initially positioned on top of another carbon atom of the graphene structure. It then migrates to the bridge site and by replacing the bridge bond it increases the size of the carbon ring (column 2 to 4). As the ring expands with the inclusion of this new carbon atom, it reaches the critical size after which it collapses.

The process of ring collapse is shown in Fig.~\ref{fig2}(c). The snapshots presented here correspond to the 40$^{th}$, 83$^{rd}$, 222$^{th}$ and 290$^{th}$ MD steps of the simulation with 42 atoms. Here the ring is composed of 14 carbon atoms before the collapse. This is just enough to form three hexagons highlighted by solid line in the fourth column of Fig.~\ref{fig2}(a). As seen in Fig.~\ref{fig2}(c), these three hexagons are consecutively formed during the collapse of the ring.

Based on the minimization of the total energy calculations using LDA (Local Density Approximation) it was predicted that a single carbon adatom is preferably adsorbed at the bridge site on graphene with a binding energy of 2.3 eV.\cite{can,can-akturk,binding} Calculations within GGA (Generalized Gradient Approximation) including van der Waals corrections and using PAW potentials also predict the bridge site as the energetically most favorable site of adsorption with a binding energy of 1.7 eV. While the energy barrier for the migration of single, isolated carbon atom is only 0.37 eV,\cite{can} it is lowered and eventually collapsed when another carbon atom is located at close proximity. Both first-principles total energy and finite temperature MD calculations have shown that initially C$_2$ and eventually C$_n$ carbon chains can form perpendicularly attached to graphene surface through inclusion of single carbon adatom one at a time. The gain of energy in the implementation of a single carbon adatom is $\sim$ 5 eV.\cite{can} However, the situation is dramatically different for a graphene sheet having armchair or zigzag edge: Single carbon atoms have shown to be attached favorably to the edge atoms with much larger binding energy (7.08 eV for armchair edge and 8.19 eV for zigzag edge). Further implementation of carbon adatoms to the edges gives rise to PH-like defect structures. Hence, the results of earlier static, zero-temperature calculations\cite{can} confirm the findings obtained from the present finite temperature MD calculations.

\subsubsection{Dimers}
\begin{figure*}
\includegraphics [width=15cm] {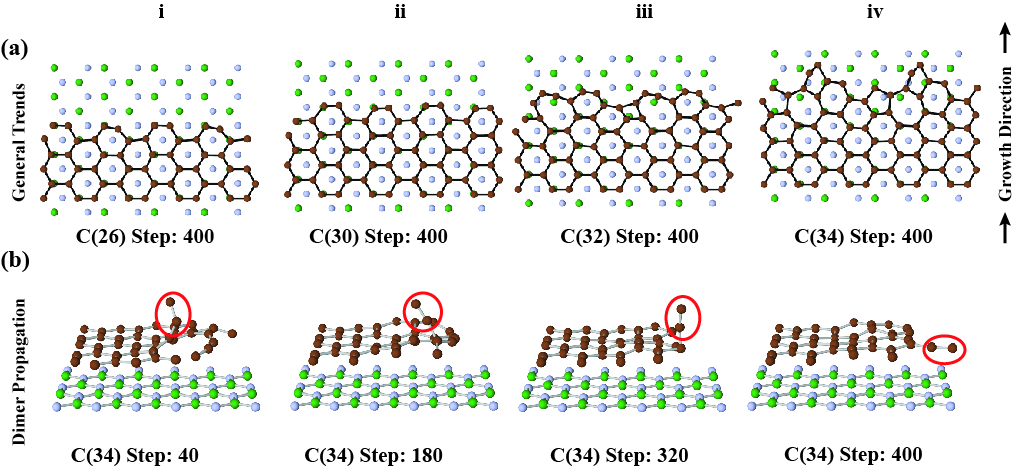}
\caption{Snapshots from ab-initio MD simulation of epitaxial growth of graphene on BN when carbon dimers are used as building blocks. B, N and C atoms are represented by green, blue and brown balls. (a) The final configurations of MD simulations involving 26, 30, 32 and 34 carbon atoms. Graphene growth is less defected as compared to growth with monomers, but ring formation and PH defects still occur as seen in columns iii and iv. (b) Migration of a carbon dimer on graphene surface. The side view snapshots are from an MD simulation having 34 carbon atoms. The dimer moves to its final position each time by binding and detaching from a different carbon atom of graphene. }

\label{fig3}
\end{figure*}

Although single carbon adatoms are the smallest building blocks of graphene, the role of carbon dimers in graphene growth was also considered.\cite{riikonen2012} The importance of carbon dimers is especially more apparent during the initial nucleation of graphene seeds due to their high mobility on certain transition metals.\cite{chen2010} In our model since we already have an initial graphene flake to which the adatoms can bind, we actually bypass the nucleation stage and directly study growth. Having studied the growth mechanisms triggered by carbon monomers, we next use carbon dimers as the building blocks. This time, we release carbon dimers from random positions on top of the graphene flake and perform molecular dynamics simulations as explained before. The snapshots taken from these simulations are presented in Fig.~\ref{fig3}. In Fig.~\ref{fig3}(a) we show the final structures obtained after 400 steps of MD simulation of 26, 30, 32 and 34 carbon atoms on BN respectively. It is seen that we have a less defected graphene growth in this case as compared to the defected structures presented in Fig.~\ref{fig2}(a). However, as seen in columns (iii) and (iv) of Fig.~\ref{fig3}(a), defects may still occur at the growing edge although less frequent as compared to growth with monomers. Just like the monomer case, we still have carbon rings forming and collapsing into PH defects as presented in Fig.~\ref{fig3}(a).

Fig.~\ref{fig3}(b) presents the migration of a carbon dimer on the graphene flake. Each of the four snapshots in this row corresponds to the 40$^{th}$, 180$^{th}$, 320$^{th}$ and 400$^{th}$ steps of the MD simulation with 34 carbon atoms. The dimer initially is bound to one of the carbon atoms on the defected graphene structure and forms a short segment of chain consisting of 3 carbon atoms. It then migrates towards the edge by bonding to another carbon atom of graphene each time and finally taking its horizontal position. Similar behaviors of long carbon chain segments on graphene surface have been previously revealed using first-principles total energy calculations.\cite{can}

\begin{figure}
\includegraphics [width=8cm] {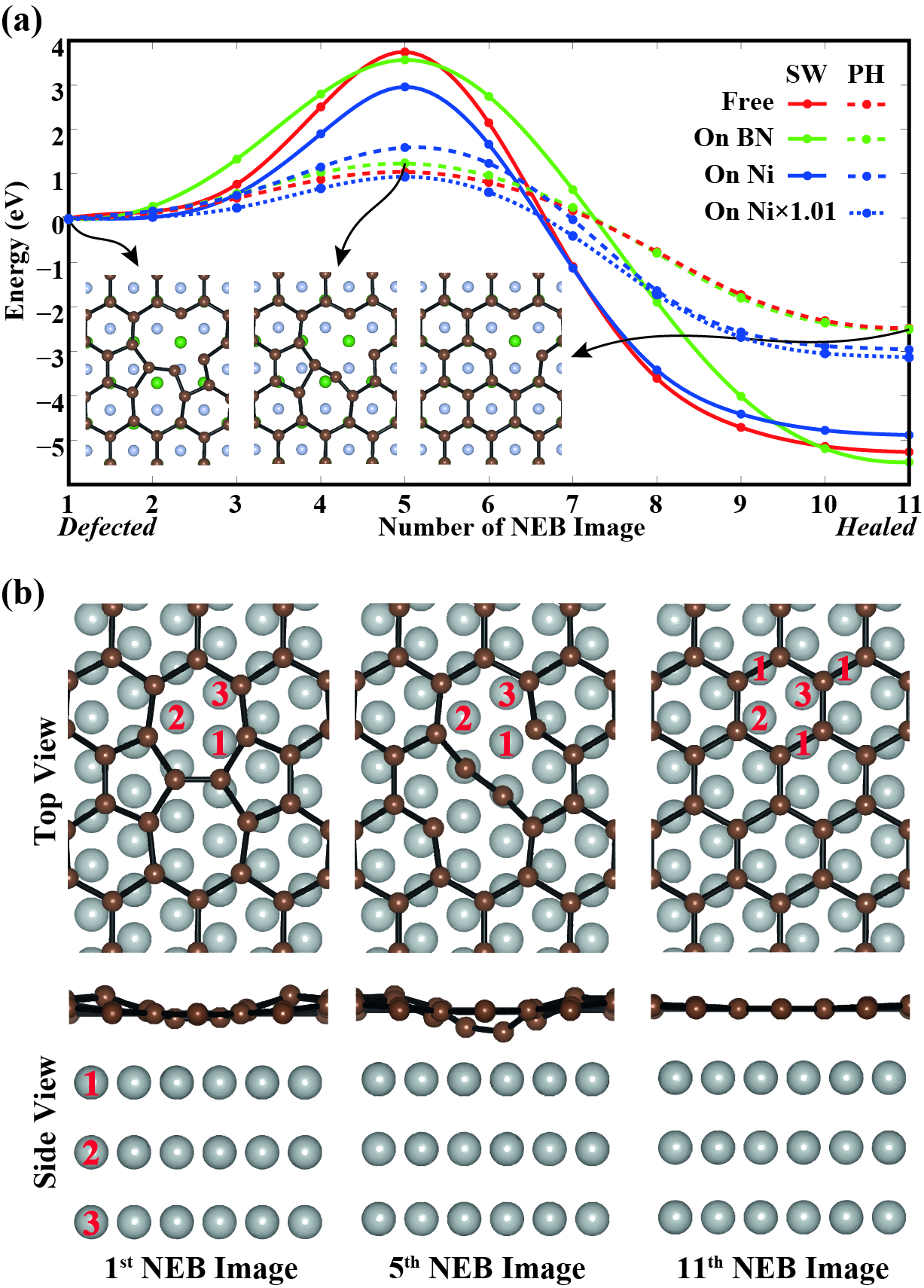}
\caption{(a) Energetics of the healing path of SW and PH defects in graphene for three cases; namely without template, graphene on BN and graphene on Ni(111) substrates. The solid red, green and blue lines show the healing path of SW defect and associated energy barriers for graphene without template, graphene on BN and graphene on Ni(111) surfaces, respectively. Energies of the defected states are set to zero. The PH healing barriers are also shown with dashed lines. The healing barrier of PH defect of graphene on Ni(111) with the lattice constant increased by 1\% is shown by dotted blue line. The inset shows the healing of PH defect of graphene on BN substrate. In the PH defect one of the carbon atoms forming the middle bond is missing the third $sp^2$-like bond. (b) Top and side views of SW defect healing on Ni substrate. The Ni atoms forming the top, middle and bottom atomic layers of the substrate are indicated by numerals 1, 2, and 3, respectively. The lateral positions of atoms in these layers are indicated by sites 1, 2, and 3. The interaction between graphene and Ni(111) is manifested in the side view of the fifth NEB image, where carbon atoms forming the C-C bond between two heptagon are pulled down when they are passing over site-2 and site-3 of the Ni substrate.}

\label{fig4}
\end{figure}

\section{Energetics of PH and SW Defect Healings}

Having noticed the role of PH defect healing in graphene growth, we move on by investigating the energetics and dynamics of SW and PH defect healing in free standing graphene, as well as graphene grown on BN and Ni(111) surfaces. The role and importance of topological SW defects on fullerene growth, isomerization and plasticity of various carbon nano structures had been previously reported. \cite{eggen96, nardelli98, samsonidze02}

Previous experimental studies have not shown any experimental evidence for the existence of SW defects in graphene. Although such defects can be observed by using tunneling electron microscopy(TEM), it was noted in previous studies that experimentally observed images of SW defects in graphene are results of electron beams which suggests that those defects are artifacts of the measurements.\cite{kotakoski2011} The presence of SW defects have also been previously studied theoretically along with the unique out of plane wavelike defect in graphene and other $sp^2$-bonded materials.\cite{ma2009} SW defects not only change the geometrical properties, but also influence electronic and chemical properties, such as band structure, reactivity and carrier transport.\cite{kang2008, carlsson2006, boukhvalov2008} In this section we calculate the energy barrier that needs to be overcame for the formation and healing of SW and PH defects. This barrier is significantly lower at the growth edges where there are vacancies nearby. Therefore, the formation and healing of these defects take place on growing edges rather than at regions where graphene has already grown to its normal structure. Hence, this explains the defect free structure of graphene once it grows successfully.

Here, we calculate the energy barrier confronted during the healing process using the climbing image NEB method.\cite{neb} Structures involved in this calculations are composed of armchair graphene nanoribbons with fixed edges and defects in the middle. The height between graphene nanoribbon edges and substrate underneath is set to the optimized value found in the case of infinite graphene sheet on infinite substrate. Also the relative position is derived in similar way. The optimum configuration of graphene on BN substrate is achieved when carbon atoms of one graphene sublattice are placed on top of boron atoms of the underlying BN layer. In case of the Ni(111) substrate, graphene structure is oriented in such a way that nickel atoms at the top layer of the substrate are under the center of the bridge bonds of graphene. We first calculate the ground state configuration of completely defected and healed states. As an initial guess of a healing path, we choose a straight line connecting these defected and healed states via linear interpolation. We choose 11 NEB images where first (defected) and eleventh (healed) are not changed while other nine images are varied until the optimum healing path is found. The fifth image was chosen as the climbing image which converges to the saddle point. Results of these calculations are outlined in Fig.~\ref{fig4}.

In Fig.~\ref{fig4}(a), the calculated healing paths of SW defect in graphene and associated barriers are shown for three cases; namely for graphene without substrate, graphene on BN and on Ni(111) surfaces. Here the energy barriers along the healing paths of SW defects are found to be 3.74~eV, 3.57~eV and 2.96~eV for free-standing, BN and Ni substrate cases, respectively. The energy barrier is significantly lowered by the substrates. The effects of substrates are proportional to their interaction energy with graphene structure. In this respect, the binding energies of graphene on BN and Ni substrates are found to be 0.13~eV and 0.41~eV per two C atoms, respectively. How the substrate can lower the barrier energy is explained by the top and side views of atomic configuration of SW defected graphene on Ni(111) substrate in Fig.~\ref{fig4}(b). Three layers of Ni(111) forming an A,B, and C stacking of the fcc structure are indicated by numerals 1, 2 and 3 starting from the top layer in the side view in Fig.~\ref{fig4} (b). Lateral positions of the atoms of these layers are indicated by sites 1, 2, and 3 in top view in the same figure. Site-2 and site-3 are energetically favorable sites for graphene atoms above Ni(111), since the binding energies of a single carbon atom on site-2 and site-3 is more favorable compared to that of site-1 by 2.48~eV and 2.46~eV, respectively. Here during the healing process of SW defect the energy barrier is lowered because of two reasons. The first reason is that, the C atoms which form the defect are pulled by site-2 and site-3 of the Ni substrate as shown in side view in Fig.~\ref{fig4}(b). This pulling is in the same direction with healing path. The second reason is that, by pulling the carbon atoms out of plane, Ni substrate increases the distance between these atoms and thereby decreases stress in the carbon-carbon bonds during the healing.

The healing paths and energy barriers of PH defect in graphene are also shown by the dotted curves in  Fig.~\ref{fig4}(a). The inset shows the atomic configuration of PH defected graphene on layered BN substrate. Unlike SW defect, here PH defect has one carbon atom with a $sp^{2}$-type dangling bond. The energy to be gained from the saturation of this dangling bond by forming a bond with a nearest C atom of the heptagon ring becomes the driving force for the healing. As a result, the energy barrier to heal the PH defect is lowered dramatically by $\sim$ 2 eV as shown in Fig.~\ref{fig4}(a). Our argument is justified by the fact that the barrier lowering occurs also for the healing of PH defect in free standing graphene. However, contrary to the effect of substrate in the healing of SW defect above, the barrier lowering effect of Ni(111) substrate is weaker than that of BN substrate. This is due to the interaction between the substrate and $sp^{2}$-type dangling bond; namely the stronger the interaction with substrate, the lower is the gain of energy upon saturation of the $sp^{2}$-type dangling bond of a two-fold coordinated carbon atom. As a result, highest barrier lowering takes place in the healing of PH defect in free standing graphene.

Finally, we simulate the effect of the the expansion of graphene lattice on the healing of defects. To this end we have performed the NEB calculation for graphene on Ni(111) surface with lattice constants increased by 1\%. It is well known that graphene has a negative thermal expansion coefficient\cite{yoon2011}, while for Ni it is positive. However, the lattice constant of graphene expands when it is stuck to Ni(111) substrate at high temperature. Under these circumstances the healing barrier of PH defect is decreased by 0.7 eV when the lattice constant is increased. This situation is shown in Fig.~\ref{fig4}(a).

\section{Conclusion}

Graphene growth and energy barrier calculations of defect healing were investigated using ab-initio MD calculations. It was found that there are two mechanisms which play crucial roles in the growth of graphene. First mechanism is the formation of large carbon rings at the edges which eventually collapse to form honeycomb structure with defects. This collapse is found to be initiated by the new coming carbon atoms which replace one of the bonds in the ring, and expands it until the critical size is reached. Second mechanism is the formation of PH defects near the edge and their healing. We have shown that the energy barrier needed to overcome during healing of the PH defects are much lower than that of the SW defects. We have shown that the presence of a BN or Ni substrate have crucial effect on growth. These substrates guide the formation of honeycomb structures from carbon rings and enable the healing of specific defects as growth proceeds. We also studied graphene growth using carbon dimers as building blocks and found that defect formation is less frequent as compared to growth with monomers.

\section{Acknowledgements}
Part of the computational resources has been provided by TUBITAK ULAKBIM, High Performance and Grid Computing Center (TR-Grid e-Infrastructure) and UYBHM at Istanbul Technical University through Grant No. 2-024-2007. This work was supported by TUBITAK and the Academy of Sciences of Turkey(TUBA).


\begin{thebibliography}{11}
\bibitem{nickel}
In the analysis of energetics of healings of SW and PH defects using the climbing image nudged elastic band (NEB) method\cite{neb} we considered the layered BN and Ni(111) substrates, while the ab-initio MD simulations of growth are performed only on layered BN substrate. Long computation times of ab-initio MD calculations using spin-polarized DFT prevented us investigating the effect of Ni(111) surface on the growth.
\end{thebibliography}
\end{document}